\documentstyle[11pt,newpasp,twoside,epsf]{article}
\def\emphasize#1{{\sl#1\/}}

\def\edcomment#1{\iffalse\marginpar{\raggedright\sl#1\/}\else\relax\fi}
\reversemarginpar

\begin{document}
\title{Recovering true metal abundances of the ICM}
 \author{S. Pellegrini}
\affil{Astronomy Department, Bologna University, 
       via Ranzani 1, I-40127 Bologna}
\author{L. Ciotti}
\affil{Bologna Astronomical Observatory, 
       via Ranzani 1, I-40127 Bologna\\
       Princeton University Observatory, 08544 Princeton, NJ, USA}

\begin{abstract}
Recovering the true {\it average} abundance of the intracluster medium
(ICM) is crucial for estimates of its global metal content, which in
turn is linked to its past evolution and to the star formation history
of the stellar component of the cluster. We analyze here how abundance
gradients affect commonly adopted estimates of the average abundance,
assuming various plausible ICM density and temperature profiles. We
find that, adopting the observed abundance gradients, the true average
mass weighted abundance is less than (although not largely deviating
from) the commonly used emission weighted abundance.
\end{abstract}

\section{Introduction}
The metal content of the ICM has proved to be a powerful tool to
constrain the past supernova history of the stellar population of
galaxy clusters, being it directly linked to the total number of
supernovae exploded in the cluster (e.g., Renzini et al. 1993).
Supernovae also provide a heating mechanism of the ICM, via galactic
winds that they power, and the relevance of this heating is a much
debated topic in recent studies of ICM evolution (e.g., Ponman,
Cannon, \& Navarro 1999; Wu, Fabian, \& Nulsen 2000; Lloyd-Davies,
Ponman, \& Cannon 2000).  So, estimates of the mass of metals in the
ICM ($M^{\rm Z}_{\rm ICM}$) are precious pieces of information, and
the more accurate they are, the more constraining they can be in
recovering the past supernova activity and from here the past
histories of star formation and heating and evolution of the ICM.  A
quantity strictly related to $M^{\rm Z}_{\rm ICM}$ is the {\it average
abundance} $<\hspace{-0.05in} Z_{\rm ICM}\hspace{-0.05in}> \equiv
M^{\rm Z}_{\rm ICM}/M_{\rm ICM}$.  There are two main possibilities to
estimate $<\hspace{-0.05in} Z_{\rm ICM}\hspace{-0.05in}>$. One is
given by:
\begin{displaymath}
\framebox{1}\hspace{0.5in}
<\hspace{-0.05in}Z_{\rm ICM}\hspace{-0.05in}>_{\rm depr} \,=\,
{\displaystyle{\int}\rho_{\rm ICM}({\bf x})\, 
                    Z_{\rm ICM}({\bf x})\, d^3{\bf x} \over 
 \displaystyle{\int}\rho_{\rm ICM}({\bf x})\, d^3{\bf x}},
\end{displaymath}
where $\rho_{\rm ICM}({\bf x})$ and $Z_{\rm ICM}({\bf x})$ are
three-dimensional gas density and abundance distributions (e.g.,
derived from deprojection of observed two-dimensional quantities).
The other is an \emphasize{emission weighted} average abundance:
\begin{displaymath}
\framebox{2}\hspace{0.5in}
<\hspace{-0.05in}Z_{\rm ICM}\hspace{-0.05in}>_{L}\,=\,
{\displaystyle{\int}n^2_{\rm ICM}({\bf x})\, 
                    \Lambda[T({\bf x}),Z({\bf x})]\, 
                    Z_{\rm ICM}({\bf x})\, d^3{\bf x} \over 
\displaystyle{\int}n^2_{\rm ICM}({\bf x})\, 
                   \Lambda[T({\bf x}),Z({\bf x})]\, d^3{\bf x}},
\end{displaymath}
where $\Lambda$ is the cooling function. This $<\hspace{-0.05in}Z_{\rm
ICM}\hspace{-0.05in}>_L$ is estimated for example as a fit parameter
when modeling the observed X-ray spectrum of the whole ICM.

\section{The problem}

There are a few serious problems with estimating
$<\hspace{-0.05in}Z_{\rm ICM}\hspace{-0.05in}>$ using method
\framebox{1}. In fact, for many/most clusters we do not know:

 -- the \emphasize{shape} of the ICM distribution;

 -- the \emphasize{viewing angles} under which we are observing the
    ICM distribution.

\par\noindent
So, we cannot unambiguously \underbar{deproject} the 
observed quantities  to derive 3-D ones such as $Z_{\rm ICM}(\bf x)$, 
$\rho_{\rm ICM}(\bf x)$ and $T_{\rm ICM}(\bf x)$. In addition, 
it is well known that deprojection is a very demanding process, sensitive 
also to: 

-- the properties of the \emphasize{instrumental PSF};

-- the \emphasize{measurement errors} (e.g., Finoguenov \& Ponman 1999).

\smallskip
\par\noindent
Correspondingly, there are clear advantages with method \framebox{2}:

 -- the knowledge of the true shape of the ICM distribution is not required
and the result is \emphasize{independent of} the viewing angles;

 -- it does \emphasize{not suffer from} instrumental effects and 
amplification of measurement errors.

 -- the average $<\hspace{-0.05in}Z_{\rm ICM}\hspace{-0.05in}>_L$ is the
\emphasize{only accessible information} that we may have for distant
clusters or those for which there are not counts enough for a
spatially resolved spectroscopy (many more of these are likely to be
observed/discovered soon, thanks to {\it Chandra} and {\it XMM}).

\smallskip\par\noindent
Unfortunately, method \framebox{2} has one problem, whose relevance
has not been investigated so far. \emphasize{Emission weighted}
average abundances $<\hspace{-0.05in}Z_{\rm ICM}\hspace{-0.05in}>_L$
derived from observed X-ray spectra are equal to the \emphasize{true}
$<\hspace{-0.05in}Z_{\rm ICM}\hspace{-0.05in}>$ only in case of a
spatially independent metal distribution. So, only if $Z({\bf x})=Z_0$
constant, then $<\hspace{-0.05in}Z_{\rm ICM}\hspace{-0.05in}>_L\, =
<\hspace{-0.05in}Z_{\rm ICM}\hspace{-0.05in}> \,= Z_0$. Recently
spatial \emphasize{abundance gradients} have been derived for the ICM
of many clusters from $ASCA$ and $BeppoSAX$ data (e.g., Finoguenov,
David \& Ponman 2000; Irwin \& Bregman 2001; De Grandi \& Molendi
2001). When such gradients are present, in general
$<\hspace{-0.05in}Z_{\rm ICM}\hspace{-0.05in}>_L\, \neq \,
<\hspace{-0.05in}Z_{\rm ICM}\hspace{-0.05in}>$, in a way dependent on
the spatial distribution of $\rho_{\rm ICM}$, $T_{\rm ICM}$, $Z_{\rm
ICM}$. Here we address the following point: how much discrepant are
$<\hspace{-0.05in}Z_{\rm ICM}\hspace{-0.05in}>_L$ and
$<\hspace{-0.05in}Z_{\rm ICM}\hspace{-0.05in}>$ ?

Our approach to find the answer is to ``calibrate'' the
$<\hspace{-0.05in}Z_{\rm ICM}\hspace{-0.05in}>_L
/<\hspace{-0.05in}Z_{\rm ICM}\hspace{-0.05in}>$ ratio using {\bf many}
plausible (spherically symmetric) profiles for $\rho_{\rm ICM}$,
$T_{\rm ICM}$ and $Z_{\rm ICM}$.  The ingredients entering the
estimates of $<\hspace{-0.05in}Z_{\rm ICM}\hspace{-0.05in}>_L$ and
$<\hspace{-0.05in}Z_{\rm ICM}\hspace{-0.05in}>$ are the following:

\par\noindent 
1) the cooling function $\Lambda=\Lambda(T,Z)$ over (0.5--10) keV;
this has been calculated with the thermal X-ray emission code of
J. Raymond (that gives the same results as the MEKAL model within
XSPEC for $kT\geq 3$ keV).

\par\noindent 
2) An abundance profile described by $Z_{\rm ICM}(r)=Z_0\times
[1-(r/r_{\rm vir})^b]$, where $b$ is a free parameter (Fig. 1) and
$r_{\rm vir}$ is the cluster virial radius.

\par\noindent 
3) Various $\rho_{\rm ICM}$ and $T_{\rm ICM}$ profiles (Figs. 2 and
3). In a {\it first set} of models we assume the ICM to be in (non
self--gravitating) hydrostatic equilibrium within a chosen cluster
(dark matter) potential well, both in the isothermal and polytropic
case.  In a {\it second set} of models we consider a cooling flow
description.

\begin{figure}
\vspace{-9truecm}
\plotone{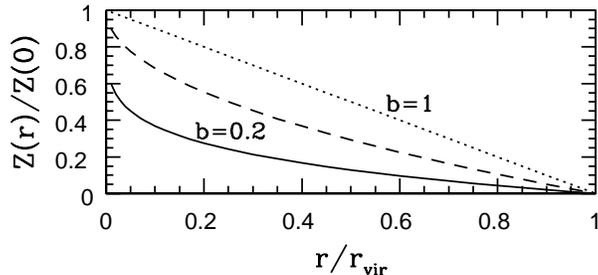}
\vspace{-1.truecm}
\caption{Abundance profiles adopted in this work}
\end{figure}

\section{Models of ICM in hydrostatic equilibrium}

\subsection{Description}
In this first set of ICM models we explored:

\smallskip
{\bf a.} The ``NFW'' cluster potential well (Navarro, Frenk \& White
1996), where the density profile of the gravitating matter is of the
form $\rho_{\rm cl}(r)=4\rho_{\rm cl}(r_s)r^3_s/r(r+r_s)^2$. The {\it
isothermal} and {\it polytropic} ICM density profiles are given by:
\begin{displaymath}
\rho_{\rm ICM}^{\rm is}(r)=\rho_0\times
\exp\displaystyle\left[{\mu m_p\over kT_0}(\phi_0-\phi)\right],
\quad
\rho_{\rm ICM}^{\rm pol}(r)=\rho_0\times 
\left[1+{\gamma-1\over\gamma}{\mu m_p\over kT_0}(\phi_0-\phi)\right]^{
{1\over \gamma -1}}.
\end{displaymath}
In the polytropic cases the temperature profile is obviously given by
\begin{displaymath}
T_{\rm ICM}^{\rm pol}(r)=T_0\times
\left[{\rho_{\rm ICM}(r)\over \rho_0}\right]^{\gamma-1}.
\end{displaymath}

\smallskip
{\bf b.} The power law models, in which the cluster
density profile is given by $\rho_{\rm cl}(r)=\rho_n\times \left(
r/r_n\right )^{-\alpha}$, with $\alpha <2$. We have now 
\begin{displaymath}
\rho_{\rm ICM}^{\rm is}(r)=\rho_0\times 
\exp\left(-{\mu m_p\over kT_0}\phi\right),\quad   
\rho_{\rm ICM}^{\rm pol}(r)=\rho_0\times
\left[1-({r\over r_{\rm vir}})^{2-\alpha}\right]^{{1\over 
\gamma -1}}.
\end{displaymath}

\smallskip
{\bf c.} The ``standard'' $\beta$--models (e.g.,
Cavaliere \& Fusco Femiano 1976), where the surface brightness profile
of the hot ICM is described by $\Sigma_{\rm X}(R)=\Sigma_{\rm X0}\times 
\left[1+\left(R/R_c\right)^2\right]^{0.5-3\beta}$, and 
$\beta=0.5-0.8$ (Mohr et al. 1999, Jones \& Forman 1999). The 
(untruncated) ICM density profiles are given by:
\begin{displaymath}
\rho_{\rm ICM}^{\rm is}(r)=\rho_0\times
\left[1+\left({r\over R_c}\right)^2\right]^{-3\beta/2},\quad
\rho_{\rm ICM}^{\rm pol}(r)=\rho_0\times
\left[1+\left({r\over R_c}\right)^2\right]^{f(\beta,\gamma)},
\end{displaymath}
where $f(\beta,\gamma)$ can be found in Ettori (2000).  Both density
distributions are then truncated at a radius $R_t=\delta R_c$, with
$\delta$ a free parameter.

Constraints on the parameter values have been derived from
observational results and from the relations $M_{200}$ -- $<T_X>$
(Navarro et al. 1996), $<T_X>$ -- $r_{200}$ (Evrard et al. 1996) and
$\sigma^2$ -- $<T_X>$ (Xue \& Wu 2000).

\smallskip
\par\noindent

\begin{figure}
\vspace{-6.5truecm}
\plotone{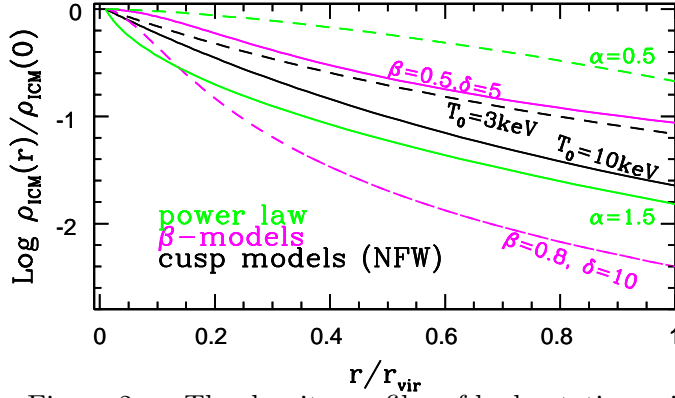}
\vspace{-1truecm}
\caption{The density profiles of hydrostatic equilibrium
models described in section 3, in the isothermal case.}
\end{figure}

\begin{figure}
\vspace{-7.5truecm}
\plotone{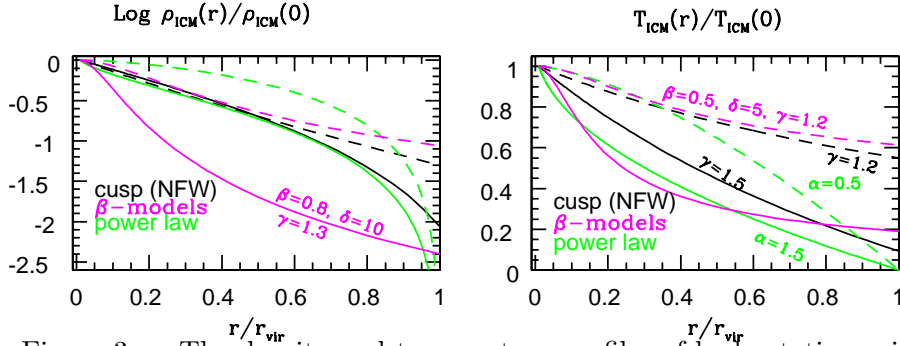}
\vspace{-0.8truecm}
\caption{The density and temperature profiles of hydrostatic equilibrium
models described in section 3, in the polytropic case.}
\end{figure}

\subsection {Results}

We estimated $<\hspace{-0.05in}Z_{\rm ICM}\hspace{-0.05in}>_L/
<\hspace{-0.05in}Z_{\rm ICM}\hspace{-0.05in}>$ by adopting the
following parameter values:

1) for the abundance profile $b=0.2$, i.e., we considered a sort of
``cuspy'' profile, as observed in some clusters (e.g., De Grandi \&
Molendi 2001; Arnaud, this meeting), and $b=1$, a linear dependence on
radius. The results are almost independent of $Z_0$.

2) the temperatures are $T_0=3$ and 10 keV in the isothermal case;
these are central temperatures, in the polytropic case.  This
assumption encompasses the values observed for rich galaxy clusters
(e.g., Markevitch et al. 1998).

3) $1< \gamma\leq 5/3$ in the polytropic case. This includes the
$\gamma=1.2$ value that gives a ``good'' fit of some observed
temperature profiles for ICM's described by a $\beta$-model
(Markevitch et al. 1998).

4) for the power law models $\alpha=0.5$ and 1.5.

5) for the $\beta$-models: $\beta=0.5$ and 0.8, $\delta=5$ and 10.

\smallskip
\par\noindent
The general trend of the values derived for 
$<\hspace{-0.05in}Z_{\rm ICM}\hspace{-0.05in}>_L/
<\hspace{-0.05in}Z_{\rm ICM}\hspace{-0.05in}>$ is that:
\smallskip
\par\noindent 
$\bullet $ $<\hspace{-0.05in}Z_{\rm ICM}\hspace{-0.05in}>_L/
<\hspace{-0.05in}Z_{\rm ICM}\hspace{-0.05in}>$ always increases for
centrally steeper abundance profiles (i.e., lower $b$ values) and
steeper gas density profiles (i.e., higher $T_0$, $\alpha$, $\beta$
and $\delta$).

\smallskip
\par\noindent
$\bullet $ $<\hspace{-0.05in}Z_{\rm ICM}\hspace{-0.05in}>_L/
<\hspace{-0.05in}Z_{\rm ICM}\hspace{-0.05in}>$ is always $>1$, and it
assumes the values below:

\smallskip
\hspace{1.5in}isothermal case \hspace{0.7in} polytropic case

\smallskip
\par\noindent
a. NFW:\hspace{1.17in} 1.4--1.7 \hspace{1.5in} 1.4--1.7

\smallskip
\par\noindent
b. POWER LAW:\hspace{0.55in} 1.3--1.8 \hspace{1.5in} 1.3--1.5

\smallskip
\par\noindent
c. $\beta$-MODELS:\hspace{0.74in} 1.4--2.1 \hspace{1.5in} 1.5--2.2 

\smallskip
\par\noindent 
$\bullet$ The typical range of values for $<\hspace{-0.05in}Z_{\rm
ICM}\hspace {-0.05in}>_L/ <\hspace{-0.05in}Z_{\rm
ICM}\hspace{-0.05in}>$ is 1.4--2, quite insensitive to variations of
density and temperature profiles in the chosen (large) range.

\section{Cooling flow models}

We assume the ICM to be described by two gas phases, at two fixed
temperatures $T_{\rm cool}$ and $T_{\rm amb}$, in pressure equilibrium
within the cooling radius $r_{\rm cool}$. Outside $r_{\rm cool}$ there
is only the ambient gas at $T_{\rm amb}$ (Ettori 2001). So:
\begin{displaymath}
\rho_{\rm ICM}(r)=\rho_{\rm 0,cool}\times \left[1-\left({r\over r_{\rm
cool}}\right)^2 \right]^{ 1.5{\beta_{\rm cool}}} + \rho_{\rm
0,amb}\times \left[1+\left({r\over {r_c}}\right)^2 \right]^{ -1.5{\beta_{\rm
amb}}}.
\end{displaymath}

\par\noindent 
Correspondingly the X-ray surface brightness profile is given by the
superposition 
\begin{displaymath}
\Sigma_{\rm X}(R)=\Sigma_{\rm X0,cool}\times
\left[1-\left({R\over r_{\rm cool}}\right)^2 \right]^{0.5+3\beta_{\rm cool}}+
\Sigma_{\rm X0,amb}\times \left[1+\left({R\over r_c}\right)^2 \right]^{
0.5-3\beta_{\rm amb}}.
\end{displaymath}

For two well studied cooling flow clusters, for which this
decomposition has been made (Ettori 2001), we find
$<\hspace{-0.05in}Z_{\rm
ICM}\hspace{-0.05in}>_L/<\hspace{-0.05in}Z_{\rm ICM}\hspace{-0.05in}>$
as given in Table 1 (for $b=0.2$ and 1).

\begin{table*}
\begin{flushleft}
\caption{Results for cooling flow models}
\begin{tabular}{ l  rl  l  l l  l  l l  l l  l }
\noalign{\smallskip}
\hline
\noalign{\smallskip}
Cluster & $\delta$ & $r_c=r_{\rm cool}$ & $T_{\rm cool}$ & $T_{\rm amb}$ & 
$\beta_{\rm cool}$ & $\beta_{\rm amb}$ &$<\hspace{-0.1in}Z_{\rm ICM}
\hspace{-0.1in}>_L/<\hspace{-0.1in}Z_{\rm ICM}\hspace{-0.1in}>$ \\
       &          & $(h_{50}^{-1}$Mpc) & (keV) & (keV) & &  &    \\
\noalign{\smallskip}
\hline
\noalign{\smallskip}
A1795 & 5.8 & 0.26 & 0.5--1.1 & 7.4 & 1.801 & 0.761 & \hspace{0.4in}1.6--2.0 \\
A2199 & 11.5 & 0.13 & 0.4--0.8 & 4.6 & 1.635 & 0.644 & \hspace{0.4in}1.7--2.1 \\
\noalign{\smallskip}
\tableline
\tableline
\end{tabular} 
\end{flushleft}
\end{table*}

\section{Future developements}

We plan to do the following further investigation:
\smallskip\par\noindent
1) Calibrate $<\hspace{-0.05in}Z_{\rm ICM}\hspace{-0.05in}>_L/
<\hspace{-0.05in}Z_{\rm ICM}\hspace{-0.05in}>$ for axisymmetric and
triaxial models (not significantly different results are expected).

\smallskip\par\noindent
2) Calibrate $<\hspace{-0.05in}Z_{\rm ICM}\hspace{-0.05in}>_{\rm
depr}/ <\hspace{-0.05in}Z_{\rm ICM}\hspace{-0.05in}>$ for axisymmetric
and triaxial models, by considering their projection at different
viewing angles, by circularizing and deprojecting their X-ray
properties, by deriving putative 3-D $\rho_{\rm ICM}(r)$ and $Z_{\rm
ICM}(r)$ and comparing $M^{\rm Z}_{\rm ICM,depr}=4\pi\int r^2\rho_{\rm
ICM}(r) Z_{\rm ICM}(r) dr$ with the true value $M^{\rm Z}_{\rm ICM}$.

\smallskip\par\noindent
3) Evaluate $<\hspace{-0.05in}Z_{\rm ICM}\hspace{-0.05in}>_L/
<\hspace{-0.05in}Z_{\rm ICM}\hspace{-0.05in}>$ for ICM's resulting
from high resolution numerical hydrodynamic simulations that include
dark matter, gas and star formation (Cen \& Ostriker 2000).

\end{document}